\documentclass[12pt,dvips]{article}

\usepackage{rotating}
\usepackage{axodraw}
\usepackage{epsfig}
\usepackage{color}

\newcommand{\imag}{\Im {\rm m}}
\newcommand{\real}{\Re {\rm e}}
\newcommand{\s}{\\ \vspace*{-3mm}}

\begin{document}

\mbox{ } \\[-1cm]
\mbox{ }\hfill DESY 02--020\\
\mbox{ }\hfill IFT--02/03\\
\mbox{ }\hfill hep--ph/0202039\\
\mbox{ }\hfill \today\\
\bigskip

\begin{center}
{\Large{\bf ANALYSIS OF THE NEUTRALINO SYSTEM\\[1mm]
             IN SUPERSYMMETRIC THEORIES\\[3mm]
             -- ADDENDUM --}} \\[2cm]
            S.Y. Choi$^1$,\, J. Kalinowski$^2$,\, 
            G. Moortgat--Pick$^3$ and P.M. Zerwas$^3$ 
\end{center}

\bigskip
\bigskip
\bigskip

{\small
\begin{enumerate}
\item[{}] $^1$ {\it Department of Physics, Chonbuk National University, Chonju
               561--756, Korea}\\[-7mm]
\item[{}] $^2$ {\it Instytut Fizyki Teoretycznej, Uniwersytet Warszawski, 
               PL--00681 Warsaw, Poland}\\[-7mm]
\item[{}] $^3$ {\it Deutsches Elektronen-Synchrotron DESY, D-22603 Hamburg, 
               Germany}
\end{enumerate}
}
\bigskip
\bigskip
\bigskip
\vskip 2cm

\begin{abstract}
\noindent
In the preceding reference \cite{CKMZ} we have shown how the 
fundamental gaugino and higgsino parameters of the chargino and neutralino
system in supersymmetric theories can be determined in high--precision
experiments at $e^+e^-$ linear colliders. Within the Minimal Supersymmetric
Standard Model these parameters can be reconstructed completely even if
only the light charginos $\tilde{\chi}^\pm_1$ and the light neutralinos 
$\tilde{\chi}^0_1$ and $\tilde{\chi}^0_2$ are kinematically accessible 
in the initial phase of these machines, as demonstrated in this
Addendum.
\end{abstract}

%


\newpage

\section{The Basis}

The fundamental parameters of the gaugino/higgsino sector in supersymmetric
theories, the U(1) and SU(2) gaugino masses $M_1$ and $M_2$, and the higgsino
mass $\mu$, can be determined very accurately in experiments at prospective
$e^+e^-$ linear colliders. This has been demonstrated in the elaborate 
analysis of Ref.\cite{CKMZ}. In the initial phase of the colliders, a total
energy of $\sqrt{s}=500\, {\rm GeV}$,  raised later to $\sim 1$ TeV, 
is planned to be reached with a high integrated luminosity of 
$\sim 1\, {\rm ab}^{-1}$ within a few years \cite{TDR}; the electron and 
positron beams are planned to be polarized with a degree of 80 and 60\%, 
respectively \cite{M-P}.\s

In many scenarios it is expected \cite{Acc} that the light 
charginos $\tilde{\chi}^\pm_1$ and the two lightest neutralinos 
$\tilde{\chi}^0_1$ and $\tilde{\chi}^0_2$ can be accessed 
kinematically\footnote{In most supergravity inspired scenarios, for example,
mass relations of the type\, 
$m_{\tilde{\chi}^\pm_1}\sim m_{\tilde{\chi}^0_2}\sim 2\, m_{\tilde{\chi}^0_1}$ 
are realized in the chargino/neutralino sector.} in the initial phase 
of the colliders. In this Addendum to Ref.\cite{CKMZ} it will be shown that in 
the Minimal Supersymmetric Standard Model (MSSM) the gaugino and higgsino 
mass parameters can be reconstructed completely in this case\footnote{For 
different strategies of determining the fundamental parameters 
see \cite{1A} and references therein.}, even in CP 
non--invariant versions of the model, by measuring the properties of 
the light particle set $\{\tilde{\chi}^\pm_1;\, \tilde{\chi}^0_1, 
\tilde{\chi}^0_2\}$.\s

In $e^+e^-$ collisions, charginos and neutralinos can be produced in 
diagonal and mixed pairs among which the reactions, giving rise to visible 
final states,  
\begin{eqnarray}
&&e^+e^-\,\rightarrow\,\,\tilde{\chi}^+_1\,\tilde{\chi}^-_1\hskip 2cm { }\\[2mm]
&&e^+e^-\,\rightarrow\,\,\tilde{\chi}^0_1\,\tilde{\chi}^0_2 
\end{eqnarray}
are of particular experimental interest in the present context.\s

In standard definition \cite{CKMZ,SONG,6A}, the diagonalization of the chargino 
matrix in the MSSM 
\begin{eqnarray}
\mbox{ }\hskip -3cm {\cal M}_C=\left(\begin{array}{cc}
                M_2                &  \sqrt{2}m_W c_\beta \\[2mm]
             \sqrt{2}m_W s_\beta  &             \mu   
                  \end{array}\right)
\end{eqnarray}
generates the light and heavy states $\tilde{\chi}^\pm_i$ ($i=1,2$), while
diagonalizing the neutralino mass matrix 
\begin{eqnarray}
\mbox{ }\hskip 2cm {\cal M}_N=\left(\begin{array}{cccc}
  M_1       &      0      &  -m_Z c_\beta s_W  & m_Z s_\beta s_W \\[2mm]
   0        &     M_2     &   m_Z c_\beta c_W  & -m_Z s_\beta c_W\\[2mm]
-m_Z c_\beta s_W & m_Z c_\beta c_W &     0    &     -\mu        \\[2mm]
 m_Z s_\beta s_W &-m_Z s_\beta c_W &  -\mu    &       0
                  \end{array}\right)\
\label{eq:massmatrix}
\end{eqnarray}
leads to four neutralino states $\tilde{\chi}^0_i$ ($i=1,2,3,4$), ordered
sequentially with rising mass. The coefficients $s_\beta =\sin\beta$, 
$c_\beta=\cos\beta$ are given by the ratio of the vacuum expectation values 
of the Higgs fields,
$\tan\beta = v_2/v_1$, and $s_W,\, c_W$ are the sine and cosine of the 
electroweak mixing angle. In CP--noninvariant theories, the mass parameters 
are complex. By reparametrization of the field basis, the SU(2) mass
parameter $M_2$ can be set real and positive, while the U(1) mass parameter
$M_1$ is assigned the phase $\Phi_1$, and the higgsino mass parameter $\mu$
the phase $\Phi_\mu$. \s

The examples presented later, will be based on a single reference point for
a CP non--invariant extension of the MSSM, compatible with all experimental
constraints \cite{EDM,WYSONG},\\[-3mm]
%
\begin{eqnarray}
{\sf RP}:\,
  \bigg(|M_1|, M_2, |\mu|;\, \Phi_1, \Phi_\mu;\, \tan\beta\bigg)
             =\bigg(100.5\,{\rm GeV},\, 190.8\, {\rm GeV},\,
                 365.1\, {\rm GeV};\,\frac{\pi}{3},\,\frac{\pi}{8};
		\, 10\bigg)
\label{eq:parameter}
\end{eqnarray}
%
These fundamental parameters generate the following light chargino and 
neutralino masses,
\begin{eqnarray}
m_{\tilde{\chi}^\pm_1}=176.0\,{\rm GeV};\qquad 
m_{\tilde{\chi}^0_1}=98.7\,{\rm GeV}\, \qquad 
m_{\tilde{\chi}^0_2}=176.3\,{\rm GeV} 
\label{eq:light chimasses}
\end{eqnarray}
while the heavy masses are given by
\begin{eqnarray}
m_{\tilde{\chi}^\pm_2}=389.3\,{\rm GeV};\qquad 
m_{\tilde{\chi}^0_3}=371.8\,{\rm GeV}\, \qquad 
m_{\tilde{\chi}^0_4}=388.2\,{\rm GeV} 
\label{eq:heavy chimasses}
\end{eqnarray}
The cross sections depend on the sneutrino and selectron masses which we
assume, for the sake of simplicity, to be measured in threshold scans\,:
\begin{eqnarray}
   m_{\tilde{\nu}_{_L}}=192.8\,{\rm GeV};\,\,   \qquad
   m_{\tilde{e}_{_L}}=208.7\,{\rm GeV}\,    \qquad
   m_{\tilde{e}_{_R}}=144.1\,{\rm GeV} 
\label{eq:semasses}
\end{eqnarray}
[Angular correlations in the production of chargino/neutralino states can be
exploited otherwise to determine the slepton masses \cite{ANGLE}.]
The cross sections for chargino and neutralino pair--production with polarized
beams are big at $\sqrt{s}=500$ GeV,
\begin{eqnarray}
&& \sigma_L\{\tilde{\chi}^{\!+}_1\tilde{\chi}^{\!-}_1\!\} = 679.5\, {\rm fb} 
   \qquad \ \
   \sigma_R\{\tilde{\chi}^{\!+}_1\tilde{\chi}^{\!-}_1\!\} = 1.04\, {\rm fb} 
   \label{eq:c11 x-section}\\[1mm]
&& \sigma_L\{\tilde{\chi}^0_1\tilde{\chi}^0_2\} = 327.9\, {\rm fb} 
   \qquad \ \
   \sigma_R\{\tilde{\chi}^0_1\tilde{\chi}^0_2\} = 16.4\, {\rm fb} 
   \label{eq:n12 x-section}
\end{eqnarray}
so that sufficiently large ensembles of events, between $\sim 7\times 10^5$ and
$1\times 10^3$ events for $\tilde{\chi}^+_1\tilde{\chi}^-_1$ and
$\tilde{\chi}^0_1\tilde{\chi}^0_2$, will be generated\footnote{Information
derived from other open channels like $\tilde{\chi}^0_2\tilde{\chi}^0_2$, etc,
can be used to refine the analysis.}, allowing the analysis of the properties 
of the chargino $\tilde{\chi}^\pm_1$ and the neutralinos $\tilde{\chi}^0_{1,2}$
at great detail. 

\section{The Chargino System}

Defining the mixing angles in the unitary matrices diagonalizing the 
chargino mass matrix ${\cal M}_C$ by $\phi_L$ and $\phi_R$ for the left--
and right--chiral fields, the fundamental SUSY parameters $M_2$, $|\mu|$,
$\cos \Phi_\mu$ and $\tan\beta$ can be derived from the chargino
masses and the cosines $c_{2L,R}=\cos 2\phi_{L,R}$ of the mixing angles,
\begin{eqnarray}
M_2&=&m_W \sqrt{\Sigma-\Delta\, (c_{2L}+c_{2R})} \label{eq:m2} \\
|\mu|&=& m_W \sqrt{\Sigma+\Delta\, (c_{2L}+c_{2R})}\label{eq:mu}\\
\cos\Phi_\mu&=&\frac{\Delta^2(2-c^2_{2L}-c^2_{2R})
                  -\Sigma}{
         \sqrt{\left[1- \Delta^2 (c_{2L}-c_{2R})^2\right]
               \left[\Sigma^2-\Delta^2(c_{2L}+c_{2R})^2\right]}}
	       \label{eq:cosphi}\\
\tan\beta&=&\sqrt{\frac{1-\Delta (c_{2L}-c_{2R})}
                       {1+\Delta (c_{2L}-c_{2R})}}\label{eq:tanb}
\end{eqnarray}
where the dimensionless quantities
\begin{eqnarray}
\Sigma = \left[m^2_{\tilde{\chi}^\pm_2}+m^2_{\tilde{\chi}^\pm_1}
              - 2 m^2_{_W}\right] /2 m^2_{_W}\qquad  {\rm and} \qquad
\Delta = \left[m^2_{\tilde{\chi}^\pm_2}-m^2_{\tilde{\chi}^\pm_1}\right]/
          4 m^2_{_W}
\end{eqnarray}
have been introduced for notational convenience. \s

If only the light charginos $\tilde{\chi}^\pm_1$ can be produced, besides
the mass $m_{\tilde{\chi}^\pm_1}$, both the mixing parameters 
$\cos 2\phi_{L,R}$ can be measured nevertheless \cite{SONG,6A}. 
The $\cos 2\phi_{L,R}$ can be determined 
uniquely if the polarized cross sections are measured at one energy 
including transverse beam polarization, or else if the longitudinally
polarized cross sections are measured at two different energies.\s

It is apparent from Eq.(\ref{eq:tanb}) that the heavy chargino mass is 
bounded from above after $m_{\tilde{\chi}^\pm_1}$ and $\cos 2\phi_{L,R}$
are measured experimentally. At the same time, it is bounded from below
by not observing the heavy chargino in mixed light$-$heavy pair production.
The ensuing constraint on the heavy chargino mass
\begin{eqnarray}
     {\textstyle{\frac{1}{2}}} \sqrt{s} - m_{\tilde{\chi}^\pm_1} 
\,\leq\, m_{\tilde{\chi}^\pm_2} 
\,\leq\, \sqrt{\, m^2_{\tilde{\chi}^\pm_1}
           +\, 4 m^2_W/| \cos 2\phi_L-\cos 2\phi_R|}
\end{eqnarray}
is quite restrictive; the upper bound can still be improved by exploiting
the slightly more restrictive, but algebraically more complicated bound 
derived from 
$|\cos\Phi_\mu|\leq 1$ in Eq.(\ref{eq:cosphi}). For the example introduced 
above, a narrow window of $324.0\, {\rm GeV} \leq m_{\tilde{\chi}^\pm_2} \leq 
389.7\, {\rm GeV}$ is predicted after initial experimentation at 
the energy $\sqrt{s}=500$ GeV.\s

If both the light chargino mass $m_{\tilde{\chi}^\pm_1}$ and the heavy
chargino mass $m_{\tilde{\chi}^\pm_2}$ can be measured, the fundamental
parameters $\{ M_2, \mu; \tan\beta\}$ can be extracted unambiguously. 
However, if $\tilde{\chi}^\pm_2$ is not accessible, it depends on the 
CP properties of the higgsino sector whether they can be determined or 
not in the chargino system alone.\s

\noindent
{\bf (A)}\, If the higgsino sector is 
     \underline{CP invariant}\footnote{Analyses of electric dipole
     moments strongly suggest that CP violation in the higgsino sector
     will be very small in the MSSM if this sector is 
     non--invariant at all \cite{EDM,WYSONG}.},
     Eq.(\ref{eq:cosphi}) can be exploited to determine 
     $m^2_{\tilde{\chi}^\pm_2}$ from $\cos\Phi_\mu =\pm 1$, up to at most 
     a two--fold ambiguity, Refs.\cite{SONG,6A}.
     This ambiguity can be resolved if other observables
     can be evaluated, {\it notabene}\,  the mixed--pair $\tilde{\chi}^0_1
     \tilde{\chi}^0_2$ production cross sections. \s

\noindent
{\bf (B)}\, If $\tilde{\chi}^\pm_2$ is not accessible,\, the parameters in
     Eqs.(\ref{eq:m2}--\ref{eq:tanb}) cannot be determined in a 
     \underline{CP non--invariant} theory in the chargino sector alone. 
     They remain dependent on the unknown heavy chargino
     mass $m_{\tilde{\chi}^\pm_2}$. Two trajectories are generated
     in $\{M_2, \mu; \tan\beta\}$ space, parametrized by 
     $m_{\tilde{\chi}^\pm_2}$ and classified by the two values $\Phi_\mu$ and
     $(2\pi -\Phi_\mu)$ for the phase of the higgsino mass parameter,
     {\it i.e.} the sign of $\sin\Phi_\mu$. 
     It will be shown in the next section that the analysis of the two
     light neutralino states $\tilde{\chi}^0_1$ and $\tilde{\chi}^0_2$ 
     can be used to predict the heavy chargino mass $m_{\tilde{\chi}^\pm_2}$
     in the MSSM. The phase ambiguity can be resolved\footnote{If not resolved,
     the two--fold ambiguity will propagate to the final set $\{M_1, M_2;
     \mu\}$ with the sign of $\imag M_1$, coupled to the sign of $\imag \mu$,
     remaining undetermined, {\it i.e.} the sign of $\sin\Phi_1$ coupled to 
     the sign of $\sin\Phi_\mu$.} by measuring the sign of CP--odd 
     observables associated with normal $\tilde{\chi}^0_2$ polarization 
     in $\tilde{\chi}^0_1 \tilde{\chi}^0_2$ pair production \cite{WYSONG}.
     Subsequently the entire set of fundamental gaugino and
     higgsino parameters can be determined uniquely.

\section{The Neutralino System}

The symmetric neutralino mass matrix ${\cal M}_N$ is diagonalized by
a unitary matrix, defined such that the mass eigenvalues 
$m_{\tilde{\chi}^0_i}$ of the four Majorana fields $\tilde{\chi}^0_i$
are positive.\s

The squared mass eigenvalues of ${\cal M}_N {\cal M}^\dagger_N$ are  
solutions of the characteristic equations \cite{CKMZ}
\begin{equation}
m_{\tilde{\chi}^0_i}^8-a\, m_{\tilde{\chi}^0_i}^6
+b\, m_{\tilde{\chi}^0_i}^4-c\, m_{\tilde{\chi}^0_i}^2+d=0 \quad  
{\rm for} \quad i=1,2,3,4
\label{eq:characteristic}
\end{equation}
with the invariants $a$, $b$, $c$ and $d$ given by the fundamental
SU(2) and U(1) gaugino mass parameters $M_2$ and $M_1$, and the higgsino
mass parameter $\mu$, {\it i.e.} the moduli $M_2$, $|M_1|$, $|\mu|$ and the
phases $\Phi_1$, $\Phi_\mu$. Each of the four invariants 
$a$, $b$, $c$ and $d$  is a binomial of $\real{M_1}=|M_1|\,\cos\Phi_1$ 
and $\imag{M_1}=|M_1|\,\sin\Phi_1$. Therefore, each of the characteristic 
equations in the set (\ref{eq:characteristic}) for the neutralino mass 
squared $m^2_{\tilde{\chi}^0_i}$ can be rewritten in the form
\begin{eqnarray}
(\real{M_1})^2+(\imag{M_1})^2+ u_i\, \real{M_1}+ v_i\, \imag{M_1}
 = w_i \quad {\rm for}\quad i=1,2,3,4
\label{eq:Mphase}
\end{eqnarray}
The coefficients $u_i$, $v_i$ and $w_i$ are functions of the 
parameters $M_2$, $|\mu|$, $\Phi_\mu$, $\tan\beta$ and the mass
eigenvalue $m^2_{\tilde{\chi}^0_i}$ for fixed $i$. The coefficient $v_i$ 
is necessarily proportional to $\sin\Phi_\mu$ because physical neutralino
masses are CP--even; the sign ambiguity for $\sin\Phi_\mu$, a result of
the two--fold cos solution $\Phi_\mu\leftrightarrow (2\pi-\Phi_\mu)$,
transfers to the associated sign ambiguity in the CP--odd quantity
$\imag M_1$, {\it i.e.} in $\sin\Phi_1$.

\begin{figure}[tbh]
\begin{center}
\vspace*{5mm}
 \epsfig{file=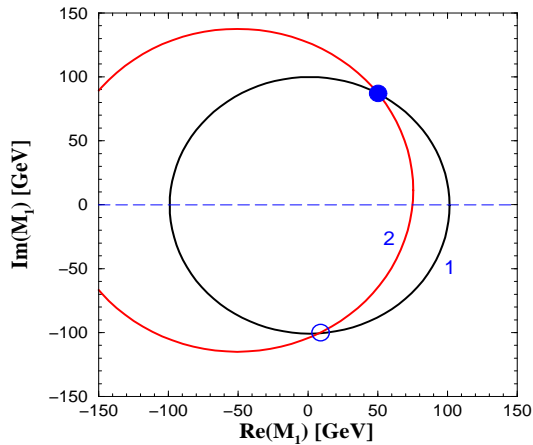,height=6cm,width=7cm}
\caption{\it The contours of two measured neutralino masses
           $m_{\tilde{\chi}^0_1}$ and $m_{\tilde{\chi}^0_2}$ in the
	   $\{{\real M_1},\,{\imag M_1}\}$ plane; the parameter set
	   $\{M_2=190.8\, {\rm GeV};\, |\mu|= 365.1\, {\rm GeV}, 
	   \Phi_\mu =\pi/8;$ $\tan\beta=10\}$ is assumed to be
	   known from the chargino sector. [The second possible solution
	   of (\ref{eq:Mphase}), with the circles reflected at the broken
	   null--line, can be rejected by measuring the sign of
	   $\sin\Phi_\mu$, related to the sign of $\sin\Phi_1$ or 
	   $\imag M_1$.]}
\label{fig:circle}
\end{center}
\end{figure}
\mbox{ }\\[-2cm]

\section{Reconstruction of the Fundamental Parameters}

The characteristic equation (\ref{eq:Mphase}) defines a circle
in the $\{\real M_1, \imag M_1\}$ plane for each neutralino mass 
$m_{\tilde{\chi}^0_i}$. With only two light neutralino masses 
$m_{\tilde{\chi}^0_1}$ and $m_{\tilde{\chi}^0_2}$ measured, we are left with
a two--fold ambiguity marked by the two, black and open,
dots in Fig.~\ref{fig:circle}.\s

\noindent
{\bf (A)}\, In the \underline{CP invariant} version of the MSSM, the 
     measurements of $m_{\tilde{\chi}^\pm_1}$ and $\cos 2\phi_{L,R}$ lead to 
     at most a two--fold ambiguity in $\{M_2, \mu; \tan\beta\}$. Inserting
     the two neutralino masses $m_{\tilde{\chi}^0_{1,2}}$ in the set
     Eq.(\ref{eq:Mphase}) for $\imag M_1=0$, this induces at most a 
     two--fold ambiguity in $M_1$. This ambiguity can finally be resolved 
     by measuring the cross sections for $\tilde{\chi}^0_1\tilde{\chi}^0_2$ 
     pair production. \s

\noindent
{\bf (B)}\, However, in scenarios with \underline{CP violation}, the 
     {\it loci} of the two 
     crossing points depend on the  unknown heavy chargino mass 
     $m_{\tilde{\chi}^\pm_2}$. The two values of $\real M_1$ and $\imag M_1$ 
     are depicted for the window of the allowed values 
     $m_{\tilde{\chi}^\pm_2}$ in the two upper panels Fig.~2a/b 
     assuming that the sign of $\sin\Phi_\mu$ will have been determined [cf.
     Footnote \# 4].\s

\begin{figure}[htb]
\begin{center}
\vspace*{0.1cm}
 \epsfig{file=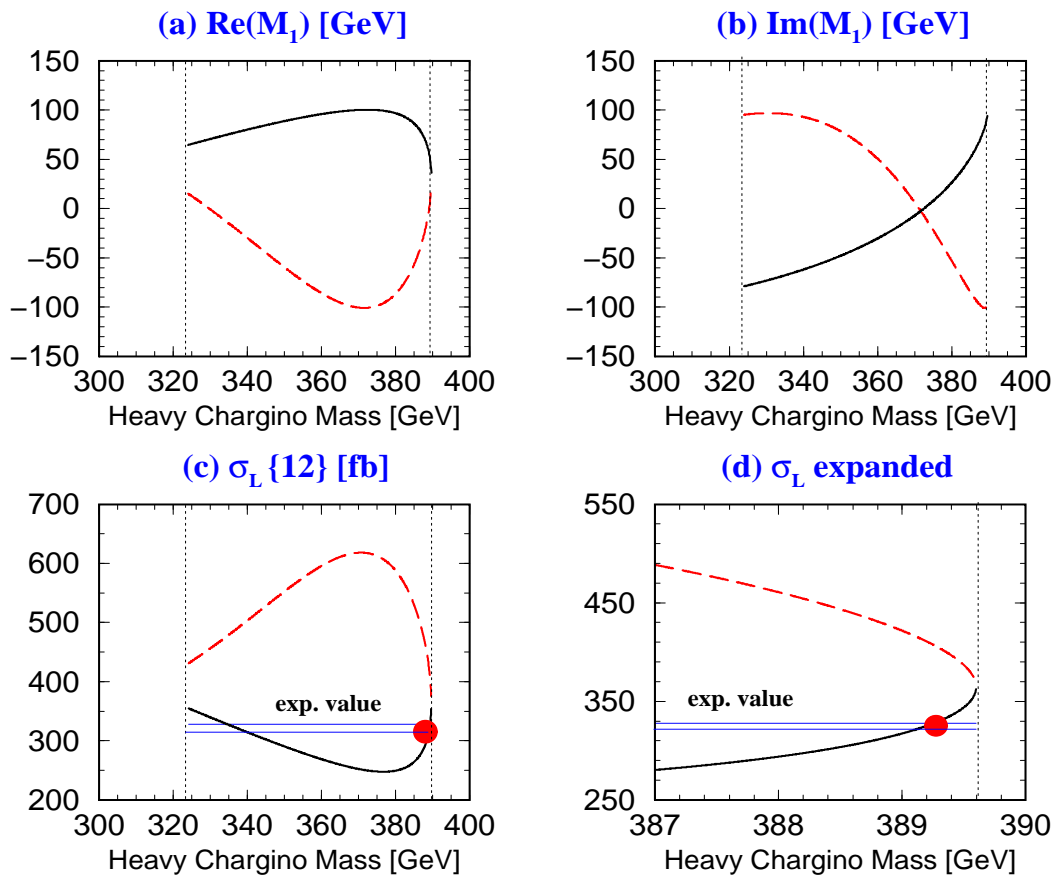,height=11.8cm,width=14cm}
\vspace*{-1mm}
\caption{\it (a,b) The two sets of $\{{\real M_1},\,{\imag M_1}\}$ denoting the 
	 crossing points of the two circles for the measured neutralino masses 
	 in Fig.\ref{fig:circle} within the allowed window; (c) the 
	 corresponding cross sections 
	 $\sigma_L\{\tilde{\chi}^0_1\tilde{\chi}^0_2\}$ as functions of the 
	 heavy chargino mass for its entire allowed mass range, and (d)
	 magnified for the unique solution $m_{\tilde{\chi}^\pm_2}=389.3$ GeV.
	 [The other solution to $\sigma_L\{\tilde{\chi}^0_1\tilde{\chi}^0_2\}$
	 can be excluded by the measurement of 
	 $\sigma_R\{\tilde{\chi}^0_1\tilde{\chi}^0_2\}$.]
	 The experimental cross section for $L$ polarization 
	 is denoted by the horizontal lines in the lower panels.}
\label{fig:m1sol}
\vspace{-0.3cm}
\end{center}
\end{figure}

     By measuring the pair--production cross sections 
     $\sigma_L\{\tilde{\chi}^0_1\tilde{\chi}^0_2\}$ and
     $\sigma_R\{\tilde{\chi}^0_1\tilde{\chi}^0_2\}$, a unique solution, 
     for both the parameters $m_{\tilde{\chi}^\pm_2}$ and $\real M_1, 
     \imag M_1$ can be found at the same time as demonstrated in the lower 
     panels Fig.~2c/d. The second solution $m_{\tilde{\chi}^\pm_2}\!\approx\!
     335$ GeV for $\sigma_L\{\tilde{\chi}^0_1\tilde{\chi}^0_2\}$ can be 
     excluded by the measurement of 
     $\sigma_R\{\tilde{\chi}^0_1\tilde{\chi}^0_2\}$ because the predicted
     value of $13.2$ fb is far away from its experimental value of
     $16.4$ fb. The cross sections depend on the neutralino mixing parameters
     which are given by the fundamental U(1) and SU(2) gaugino and higgsino
     parameters. They are parametrized therefore solely by
     $m_{\tilde{\chi}^\pm_2}$ after the chargino system and the two neutralino 
     masses $m_{\tilde{\chi}^0_{1,2}}$ are evaluated as elaborated before. 
     As a result, the additional measurement of the cross sections leads to a 
     unique solution for $m_{\tilde{\chi}^\pm_2}$ and subsequently to a unique 
     solution for $\{M_1, M_2; \mu; \tan\beta\}$ [assuming that the discrete 
     CP ambiguity in the associated signs of $\sin\Phi_\mu$ and 
     $\sin\Phi_1$ has been resolved by measuring the normal $\tilde{\chi}^0_2$ 
     polarization].\s\s
     
\begin{figure}[hbt]
\begin{center}
\vspace*{2mm}
 \epsfig{file=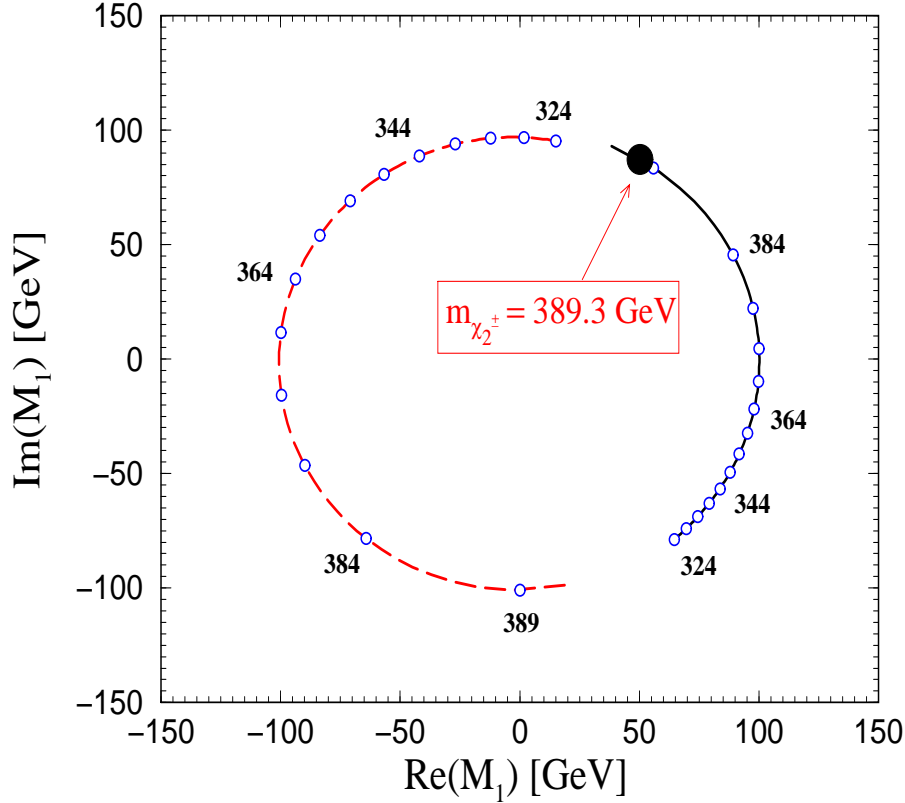,height=11cm,width=12cm}
\caption{\it The two trajectories of the crossing points of the two circles for
           the masses $m_{\tilde{\chi}^0_1}$ and $m_{\tilde{\chi}^0_2}$ in the
	   $\{{\real M_1},\,{\imag M_1}\}$ plane, parameterized by the heavy
	   chargino mass $m_{\tilde{\chi}^\pm_2}$. The small open 
	   circles denote the heavy chargino mass parameter spaced by 5 GeV; 
	   the unique solution which is determined by the measurement of the 
	   pair--production cross sections, is marked by the black dot.}
\label{fig:m1sol}
\end{center}
\end{figure}
\vskip -0.2cm

     This procedure can nicely be summarized in a single figure: Fig.~3.
     The two crossing points of the masses $m_{\tilde{\chi}^0_1}$ and
     $m_{\tilde{\chi}^0_2}$ in Fig.\ref{fig:circle} define the two 
     trajectories in the complex
     $M_1$ plane, parametrized by the heavy chargino mass 
     $m_{\tilde{\chi}^\pm_2}$. The $L$ and $R$ cross sections vary along
     the two trajectories; comparing the predicted values with the measured
     values leads to a unique solution on the trajectories marked by a
     black dot, {\it i.e.} to unique values for $M_1$ and 
     $m_{\tilde{\chi}^\pm_2}$, along with finally unique solutions
     for $M_2, \mu$ and $\tan\beta$.\\ 

{\it To summarize.} If only the light chargino $\tilde{\chi}^\pm_1$
and the two light neutralinos $\tilde{\chi}^0_1$ and $\tilde{\chi}^0_2$
can be accessed kinematically in the initial phase of $e^+e^-$ linear 
colliders, measurements of the masses $m_{\tilde{\chi}^0_1}$ and
$m_{\tilde{\chi}^0_2}$, and the neutralino production cross section
and $\tilde{\chi}^0_2$ polarization in the process 
$e^+e^-\rightarrow \tilde{\chi}^0_1 \tilde{\chi}^0_2$, combined with the light 
chargino mass $m_{\tilde{\chi}^\pm_1}$ and the chargino production cross 
section of the process $e^+e^-\rightarrow \tilde{\chi}^+_1 \tilde{\chi}^-_1$ 
for polarized beams, allow us to perform a complete and precise analysis of
the basic MSSM parameters in the gaugino/higgsino sector: $\{M_1, M_2; \mu;
\tan\beta\}$.\s\s

\subsection*{Acknowledgments}

SYC is supported by the Korea Science and Engineering Foundation
(KOSEF) through the KOSEF--DFG collaboration project, Project No.
20015--111--02--2, JK by the KBN Grant No. 
2P03B 060 18. The work is supported in part by the European Commission 5-th
Framework Contract HPRN-CT-2000-00149.\s

\end{document}